\documentclass[prd,twocolumn,tightenlines,superscriptaddress,nofootinbib,showpacs]{revtex4}
\usepackage{amssymb,latexsym}
\usepackage{amsmath,amsbsy,bbm}
\usepackage{epsfig,bm}
\usepackage{graphicx,comment}
\usepackage{color}
\newcommand{\p}{\partial}

\begin{document} 

\title{Monte Carlo Determination of the Low-Energy Constants 
of a Spin 1/2 Heisenberg Model with Spatial Anisotropy}

\author{F.-J.~Jiang}
\email[]{fjjiang@itp.unibe.ch}
\affiliation{Center for Research and Education in Fundamental Physics, 
Institute for Theoretical Physics, Bern University, 
Sidlerstrasse 5, CH-3012 Bern, Switzerland}
\author{F.~K\"ampfer}
\affiliation{Condensed Matter Theory Group, Department of Physics, 
Massachusetts Institute of Technology (MIT), 77 Massachusetts Avenue,
Cambridge, MA 02139, U.S.A.}

\author{M.~Nyfeler}
\affiliation{Center for Research and Education in Fundamental Physics, 
Institute for Theoretical Physics, Bern University, 
Sidlerstrasse 5, CH-3012 Bern, Switzerland}

\begin{abstract}

Motivated by the possible mechanism for the pinning of the electronic liquid
crystal direction in YBCO as proposed in \cite{Pardini08}, we use the 
first principles Monte Carlo method to study the spin 1/2 Heisenberg model with
antiferromagnetic couplings $J_{1}$ and $J_{2}$ on the square lattice. 
In particular, the low-energy constants spin stiffness $\rho_s$, staggered 
magnitization ${\cal M}_s$ and spin wave velocity $c$ are determined
by fitting the Monte Carlo data to the predictions of magnon chiral
perturbation theory.
Further, the spin stiffnesses $\rho_{s1}$ and $\rho_{s2}$ as a 
function of the ratio $J_{2}/J_{1}$ of the couplings are investigated in detail.
Although we find a good agreement between our results with those 
obtained by the series expansion method in the weakly anisotropic regime,
for strong anisotropy we observe discrepancies.

\end{abstract}

\maketitle

{\bf Introduction}.---
Understanding the mechanism responsible for high-temperature 
superconductivity in cuprate materials remains one of the most active 
research fields in condensed matter physics. Unfortunately, the 
theoretical understanding of the high-$T_c$ materials using analytic 
methods as well as first principles Monte Carlo simulations is hindered
by the strong electron correlations in these materials. Despite 
this difficulty, much effort has been devoted to investigating the properties 
of the relevant $t$-$J$-type models for the 
high-$T_c$ cuprates \cite{Eder96,Lee97,Hamer98,Brunner00}. 
Although a conclusive agreement regarding the mechanism
responsible for the high-$T_c$ phenomena has not been reached yet, 
it is known that the high-$T_c$ cuprate superconductors
are obtained by doping the antiferromagnetic insulators with
charge carriers. This has triggered vigorous studies of undoped
and lightly doped antiferromagnets. Today, the undoped antiferromagnets on 
the square lattice such as La$_2$CuO$_4$ are among the quantitatively best
understood condensed matter systems. 

Spatially anisotropic Heisenberg models have been studied intensely 
due to their phenomenological importance as well as from the perspective
of theoretical interest \cite{Parola93,Sandvik99,Irkhin00,Kim00}.
For example, numerical evidence indicates that the anisotropic
Heisenberg model with staggered arrangement of the antiferromagnetic 
couplings may belong to a new universality class, in contradiction
to the $O(3)$ universality predictions \cite{Wenzel08}.  
Further, it is argued that the Heisenberg model with spatially
anisotropic couplings 
$J_{1}$ and $J_{2}$ is relevant to the newly discovered pinning effects of 
the electronic liquid crystal in the underdoped cuprate superconductor 
YBa$_2$Cu$_3$O$_{6.45}$ \cite{Hinkov2007,Hinkov2008}. It is observed that the
YBa$_2$Cu$_3$O$_{6.45}$ compound has a tiny in-plane lattice anisotropy
which is strong enough to pin the orientation of the electronic liquid
crystal in a particular direction. The authors of \cite{Pardini08} 
demonstrated that the in-plane anisotropy of the spin stiffness of the 
Heisenberg model with spatially anisotropic couplings $J_{1}$ and $J_{2}$ 
can provide a possible mechanism for the pinning of the electronic liquid 
crystal direction in YBa$_2$Cu$_3$O$_{6.45}$.

Since the anisotropy of the spin stiffness in the spin 1/2 Heisenberg model
with different antiferromagnetic couplings $J_{1}$ and $J_{2}$ has not been 
studied in detail before
with first principles Monte Carlo methods, in this letter we perform a 
Monte Carlo calculation to determine the low-energy constants, namely the 
spin stiffnesses $\rho_{s1}$ and $\rho_{s2}$, staggered magnitization
${\cal M}_s$ and spin wave velocity $c$. In particular, 
we investigate the $J_{2}/J_{1}$-dependence 
of $\rho_{s1}$ and $\rho_{s2}$, and find good 
agreement with earlier studies \cite{Pardini08} using series expansion 
methods in the weakly anisotropic regime. 
Our this finding would lead to very strong pinning energy per Cu site in
YBa$_2$Cu$_3$O$_{6.45}$ as claimed in \cite{Pardini08}.
However, deviations appear as one
moves toward strong anisotropy. We argue that the deviations observed 
between our results and the naive expectation might indicate  
an unexpected behavior of the spin-stiffness $\rho_s$ at extremely 
strong anisotropy.

\begin{figure}
\begin{center}
\includegraphics[width=0.32\textwidth]{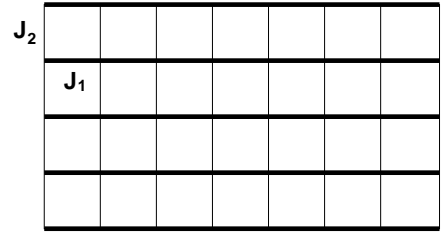}
\end{center}\vskip-0.5cm
\caption{The anisotropic Heisenberg model investigated in this study.
$J_1$ and $J_2$ are the antiferromagnetic couplings in the $1$- and 
$2$-directions, respectively.}
\label{fig0}
\end{figure}

{\bf Microscopic Models and Corresponding Observables}.---
The Heisenberg model we consider in this study is defined by the Hamilton 
operator
\begin{eqnarray}
\label{hamilton}
H = \sum_{x} \Big[\,J_{1}\vec S_x \cdot \vec S_{x+\hat{1}}+J_{2} 
\vec S_x \cdot \vec S_{x+\hat{2}}\,\Big],
\end{eqnarray}
where $\hat{1}$ and $\hat{2}$ refer to the two spatial unit-vectors. Further, 
$J_{1}$ and $J_{2}$ in eq.~(\ref{hamilton}) are the antiferromagnetic couplings 
in the
$1$- and $2$-directions respectively.
A physical quantity of central interest is the 
staggered susceptibility 
(corresponding to the third component of the staggered magnetization $M_s^3$) 
which is given by
\begin{eqnarray}
\label{defstagg}
\chi_s 
&=& \frac{1}{L_{1} L_{2}} \int_0^\beta dt \ \frac{1}{Z} 
\mbox{Tr}[M^3_s(0) M^3_s(t) \exp(- \beta H)].
\end{eqnarray}
Here $\beta$ is the inverse temperature, $L_{1}$ and $L_{2}$ are the spatial box
sizes in the $1$- and $2$-direction, respectively, and 
$Z = \mbox{Tr}\exp(- \beta H)$
is the partition function. The staggered magnetization order parameter 
$\vec{M}_s$ is defined as $\vec M_s = \sum_x (-1)^{x_1+x_2} \vec S_x$.
Another relevant quantity is the uniform susceptibility which is 
given by
\begin{eqnarray}
\label{defuniform}
\chi_u 
&=& \frac{1}{L_{1} L_{2}} \int_0^\beta dt \ \frac{1}{Z} \mbox{Tr}[M^3(0) M^3(t)
\exp(- \beta H)].
\end{eqnarray}
Here $\vec{M} = \sum_x \vec S_x$ 
is the uniform magnetization. Both $\chi_s$ and $\chi_u$ can be measured very 
efficiently with the loop-cluster algorithm using improved estimators 
\cite{Wie94}. In particular, in the multi-cluster version of the algorithm the 
staggered susceptibility is given in terms of the cluster sizes $|{\cal C}|$ 
(which have the dimension of time), i.e.
$\chi_s = \frac{1}{ \beta L_{1}L_{2}} \left\langle \sum_{\cal C} |{\cal C}|^2 
\right\rangle$.
Similarly, the uniform susceptibility
$\chi_u = \frac{\beta}{ L_{1}L_{2}} \left\langle W_t^2 \right\rangle =
\frac{\beta}{ L_{1}L_{2}} \left\langle \sum_{\cal C} W_t({\cal C})^2 
\right\rangle$
is given in terms of the temporal winding number 
$W_t = \sum_{\cal C} W_t({\cal C})$ which is the sum of winding numbers
$W_t({\cal C})$ of the loop-clusters ${\cal C}$ around the Euclidean time 
direction. Similarly, the spatial winding numbers are defined by 
$W_i = \sum_{\cal C} W_i({\cal C})$ with $i \in \{1,2\}$.

{\bf Low-Energy Effective Theory for Magnons}.---
Due to the spontaneous breaking of the $SU(2)_s$ spin symmetry down to its 
$U(1)_s$ subgroup, the low-energy physics of antiferromagnets is governed by
two massless Goldstone bosons, the antiferromagnetic spin waves or magnons. The
description of the low-energy magnon physics by an effective theory was 
pioneered by Chakravarty, Halperin, and Nelson in \cite{Cha89}.  
A systematic 
low-energy effective field theory for magnons was further developed in 
\cite{Neu89,Has90,Has91}. The staggered magnetization of an 
antiferromagnet is described by a unit-vector field $\vec{e}(x)$ in the 
coset space $SU(2)_s/U(1)_s = S^2$, i.e. 
$\vec e(x) = \big(e_1(x),e_2(x),e_3(x)\big)$ with $\vec e(x)^2 = 1$.
Here $x = (x_1,x_2,t)$ denotes a point in (2+1)-dimensional space-time. To 
leading order, the Euclidean magnon low-energy effective action takes the form
\begin{eqnarray}
\label{action}
S[\vec e\,] &=& \int^{L_1}_{0} dx_1 \int^{L_2}_{0} dx_2 \int^{\beta}_{0} 
\ dt \  
\left(\frac{\rho_{s1}}{2}\p_1 \vec e \cdot \p_1 \vec e \right. \nonumber \\
&&+\left. \frac{\rho_{s2}}{2}\p_2 \vec e \cdot \p_2 \vec e + 
\frac{\rho_s}{2c^2} \p_t \vec e \cdot \p_t \vec e\right),
\end{eqnarray}
where the index $i \in \{1,2\}$ labels the two spatial directions and $t$ 
refers to the Euclidean time-direction. The parameters $\rho_s=\sqrt{\rho_{s1}\rho_{s2}}$, $\rho_{s1}$ 
and $\rho_{s2}$ are the spin stiffness in the temporal and 
spatial directions, respectively, and $c$ is the spin wave velocity.
Rescaling $x'_{1} = (\rho_{s2}/\rho_{s1})^{1/4} x_{1}$ 
and $x'_{2} = (\rho_{s1}/\rho_{s2})^{1/4} x_{2}$, eq.~(\ref{action}) can be 
rewritten as
\begin{eqnarray}
\label{action1}
S[\vec e\,] &=& \int^{L'_{1}}_{0} dx'_1 \int^{L'_{2}}_{0} dx'_2 
\int^{\beta}_{0} dt \  
\frac{\rho_s}{2}\Big(\p'_i \vec{e} \cdot \p'_i \vec{e} \nonumber \\ 
&& + \frac{1}{c^2} \p_t \vec{e} \cdot \p_t \vec{e}\Big).
\end{eqnarray}
Additionally requiring
$L'_{1} = L'_{2} = L$ we obey the condition of square area.
Notice the effective field theories described by eqs.~(\ref{action}) and 
(\ref{action1}) are valid as long as the conditions $L_{i} \beta \rho_{s1} \gg 1$ 
and $L_{i} \beta \rho_{s2} \gg 1$ for $i\in \{1,2\}$ hold, which is indeed the case 
for the set up of this study.  
Once these conditions are satisfied, the low-energy physics of the underlying
microscopic model can be captured quantitatively 
by the effective field theory as demonstrated in \cite{Wie94}.
Further, in the so-called cubical regime (to be defined later)
which is relevant to our study,
the cut-off effects appear in the free energy density only at 
next-to-next-to-next-to-leading order (NNNLO).
The finite cut-off leads to higher-order terms in the effective Lagrangian 
due to the breaking of some symmetries and it introduces the cut-off 
dependence in the Fourier 
integrals (sums). By employing similar arguments as those presented 
in \cite{Has93}, one can show that higher-order corrections to 
eq. (\ref{action}) contain four derivatives and the leading 
cut-off effect in the Fourier integrals (sums) enters the free 
energy density
only at NNNLO. Therefore eq.~(\ref{action1}) is 
sufficient to derive up to next-to-next-to-leading order (NNLO) contributions
to the observables considered here. We have further verified that the 
inclusion 
of NNNLO contributions to the
relevant observables considered here lead to 
statistically consistent results with those not taking such corrections
into account. Hence the volume- and temperature- 
dependence of $\chi_s$ and $\chi_u$ up to NNLO (to be presented below)
are sufficient to describe our numerical data quantitatively, and
the finite cut-off effects are negligible.
Using the above Euclidean action (\ref{action1}), 
detailed calculations of a variety of 
physical quantities including the NNLO contributions 
have been carried out in \cite{Has93}. Here we only quote the results that are 
relevant to our study, namely the finite-temperature and finite-volume effects
of the staggered susceptibility and the uniform 
susceptibility. The aspect ratio of a 
spatially quadratic space-time box with box size $L$ is characterized 
by $l = (\beta c /L )^{1/3}\,,$ with which one distinguishes cubical space-time
volumes with $\beta c \approx L$ from cylindrical ones with $\beta c \gg L$. 
In the cubical regime, the volume- and temperature-dependence of the staggered 
susceptibility is given by
\begin{eqnarray}
\label{chiscube}
\chi_s &=& \frac{{\cal M}_s^2 L^2 \beta}{3} 
\left\{1 + 2 \frac{c}{\rho_s L l} \beta_1(l) \right. \nonumber \\
&&+\left.\left(\frac{c}{\rho_s L l}\right)^2 \left[\beta_1(l)^2 + 
3 \beta_2(l)\right] + O\left(\frac{1}{L^3}\right) \right\},
\end{eqnarray}
where ${\cal M}_s$ is the staggered magnetization density. Finally 
the uniform susceptibility takes the form
\begin{eqnarray}
\label{chiucube}
\chi_u &=& \frac{2 \rho_s}{3 c^2} 
\left\{1 + \frac{1}{3} \frac{c}{\rho_s L l} \widetilde\beta_1(l) +
\frac{1}{3} \left(\frac{c}{\rho_s L l}\right)^2 \times \right. \nonumber \\
&&\left.
\left[\widetilde\beta_2(l) - \frac{1}{3} \widetilde\beta_1(l)^2 - 6 \psi(l)
\right]
+ O\left(\frac{1}{L^3}\right) \right\}.
\end{eqnarray}
In (\ref{chiscube}) and (\ref{chiucube}), the 
functions $\beta_i(l)$, $\widetilde\beta_i(l)$, and $\psi(l)$, which only 
depend on $l$, are shape coefficients of the space-time box defined in 
\cite{Has93}.

\begin{figure}
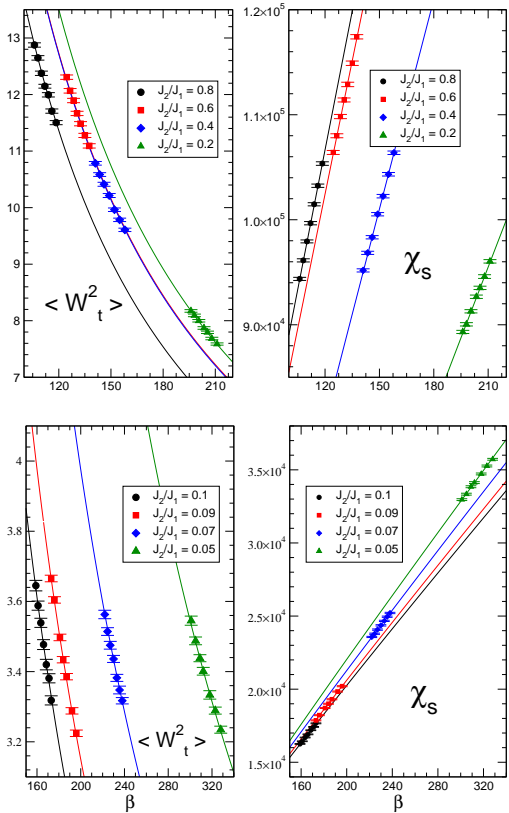

\begin{center}
\vbox{
\includegraphics[width=0.3725\textwidth]{chi_su_1.eps}
\includegraphics[width=0.3725\textwidth]{chi_su_2.eps}
}
\end{center}\vskip-0.5cm
\caption{Comparison between our numerical results (data points)
and the theoretical predictions (solid lines) which are
obtained by using the low-energy parameters from the fits.}
\label{fig1}
\end{figure}

\begin{figure}
\begin{center}
\includegraphics[width=0.375\textwidth]{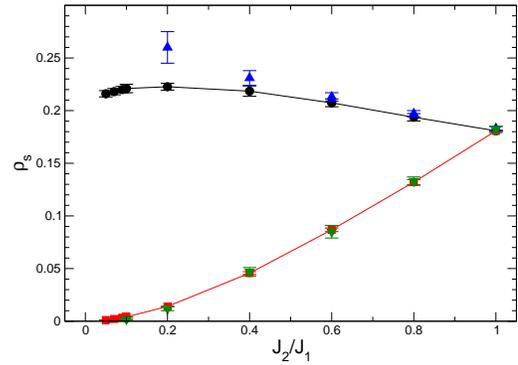}
\end{center}\vskip-0.5cm
\caption{The $J_{2}/J_{1}$-dependence of the spin stiffness $\rho_{s1}$ and
$\rho_{s2}$ of the anisotropic Heisenberg model. While the solid  
circles (black) and squares (red) are the Monte Carlo results 
of $\rho_{s1}$ and $\rho_{s2}$, respectively, the up and down triangles are
the series expansion results of \cite{Pardini08} for $\rho_{s1}$ 
and $\rho_{s2}$, respectively. The solid lines are added to guide
the eye.}
\label{fig2}
\end{figure}

\begin{figure}
\begin{center}
\includegraphics[width=0.3725\textwidth]{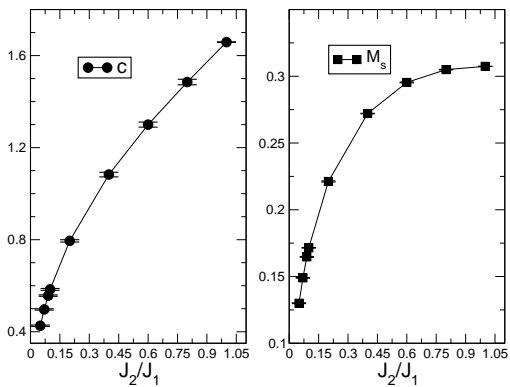}
\end{center}\vskip-0.5cm
\caption{The $J_{2}/J_{1}$-dependence of the spin wave velocity $c$ (left) and 
the staggered magnitization density ${\cal M}_s$ (right) of the anisotropic 
Heisenberg model. The solid lines are added to 
guide the eye.}
\label{fig3}
\end{figure}

{\bf Determination of the Low-Energy Parameters and 
Discussions}.---
In order to determine the low-energy constants for the anisotropic Heisenberg
model given in (\ref{hamilton}), we have performed simulations within the 
range 
$ 0.05 \leq J_{2}/J_{1} \leq 1.0$. The cubical regime is determined by
the condition $\,\langle \sum_C W_{1}(C)^2 \,\rangle \approx \langle\, 
\sum_C W_{2}(C)^2 \, \rangle  \approx \langle\, \sum_C W_{t}(C)^2\, \rangle$ 
(which implies $\beta c \approx L$). Notice that since $J_{2} \leq J_{1}$ 
in our simulations,
one must increase the lattice size $L_{1}$ in order to fulfill the condition
$\,\langle \sum_C W_{1}(C)^2 \,\rangle = \langle\, 
\sum_C W_{2}(C)^2 \,\rangle$ 
because eqs.~(\ref{chiscube}) and (\ref{chiucube}) are obtained for a
(2~+~1)-dimensional box with equal extent in the two spatial directions.
Therefore, an interpolation of the data points is required in order to
be able to use eqs.~(\ref{chiscube}) and (\ref{chiucube}). 
Further, the low-energy parameters are extracted by fitting the Monte Carlo data to 
the effective field theory predictions. 
The quality of these fits is good as can be seen from figure \ref{fig1} 
(the $\chi^2/{\text{d.o.f}}$ for all the fits 
is less than $1.25$). Figure \ref{fig2} shows 
$\rho_{s1}$ and $\rho_{s2}$, obtained from the fits, as
functions of the ratio of the antiferromagnetic couplings, 
$J_{2}/J_{1}$. 
The values of $\rho_{s1}$ ($\rho_{s2}$) obtained here
agree quantitatively with those obtained using the series expansion
in \cite{Pardini08} at $J_{2}/J_{1} = 0.8$ and $0.6$ 
($0.8$, $0.6$, $0.4$, and $0.2$). At $J_2/J_1 = 0.4$, the value we obtained
for $\rho_{s1}$ is only slightly below the corresponding series expansion 
result in \cite{Pardini08}. 
However, sizable deviations begin to 
show up for stronger anisotropies. 
Further, we have not observed the saturation of $\rho_{s1}$ to a $1$-D
limit, namely $0.25J_{1}$ as suggested in \cite{Pardini08}, even at 
$J_{2}/J_{1}$ as small as $0.05$. In 
particular, $\rho_{s1}$ decreases slightly as one moves from
$J_{2}/J_{1} = 0.1$ to $J_{2}/J_{1} = 0.05$, although they still agree
within statistical errors. Of course, one cannot rule out 
that the anisotropies
in $J_{2}/J_{1}$ considered here are still too far away from the 
regime where this particular
Heisenberg model can be effectively described by its $1$-D limit.
On the other hand, the Heisenberg model considered here and its 1-D limit
are two completely different systems, 
because spontaneous symmetry breaking appears only in 2-D, still
$\xi=\infty$ in both cases.
Further, the low-temperature behavior of $\chi_u$  in the 1-D system 
is known to be completely different from that of the 2-D system 
\cite{Egg94,Has93}.
Although intuitively one might expect a 
continuous transition of $\rho_{s1}$, 
one cannot rule out an unexpected  behavior of $\rho_{s1}$
as one moves from this Heisenberg model toward its 1-D limit. 
In particular, since earlier studies indicate that long-range order already 
sets in even for infinitesimal small $J_{2}/J_{1}$ 
\cite{Sandvik99,Affleck94,Miy95},
it would be interesting to consider even stronger anisotropies
$J_{2}/J_{1}$ than
those used in this study to see how $\rho_{s1}$ approaches its $1$-D 
limit.
In addition to $\rho_{s1}$ and $\rho_{s2}$, we have obtained
${\cal M}_s$ and $c$ as functions of $J_{2}/J_{1}$ as well from the 
fits (figure \ref{fig3}).
The values we obtained for ${\cal M}_s$ agree with earlier results
in \cite{Sandvik99}, but have much smaller errors at strong anisotropies.

Next, we would like to turn to discussing the relevance of our results to
the pinning effect observed empirically in YBa$_2$Cu$_3$O$_{6.45}$.
In \cite{Pardini08} it is argued that the $J_{2}/J_{1}$ dependence of
the spin-stiffnesses in the spatially anisotropic Heisenberg model 
studied in this work would 
lead to a very strong pinning energy per Cu site 
(one order of magnitude larger compared to the corresponding 
pinning energy in La$_2$CuO$_4$).   
To be more precise, it is the quantity $\kappa$ which is defined by 
$\rho_{s2}/\rho_{s1} = 1 + \kappa (J_{2}/J_{1} - 1)$ in the weak anisotropy
regime that results in the claim made in \cite{Pardini08}. Since
the spin-stiffnesses calculated here agree with those obtained
by series expansion in the weak anisotropy regime, which in turn implies
that our $\kappa$ agrees with that in \cite{Pardini08}, we conclude that 
the pinning energy per Cu site is indeed very strong. 
Hence the in-plane anisotropy of the spin stiffness of 
the Heisenberg model with anisotropic 
couplings $J_{1}$ and $J_{2}$ can indeed provide a possible mechanism 
for the pinning 
of the electronic liquid crystal direction in YBa$_2$Cu$_3$O$_{6.45}$.

{\bf Conclusions}.---
In this note, we have numerically studied the Heisenberg model with 
anisotropic couplings $J_{1}$ and $J_{2}$ using a loop cluster algorithm. 
The coresponding low-energy constants are determined with high precision.
Further, the $J_{2}/J_{1}$-dependence of
$\rho_{s1}$ and $\rho_{s2}$ is investigated in detail and our results 
agree quantitatively with those obtained by series expansion \cite{Pardini08}
in the weakly anisotropic regime. On the other hand, we observe 
discrepancies between our results and series expansion results in the 
strongly anisotropic regime. However, the results of our study still 
lead to very strong pinning energy per Cu site in YBa$_2$Cu$_3$O$_{6.45}$ 
which agrees with the 
claim made by the authors in \cite{Pardini08}. Finally 
we find that an unexpected behavior of $\rho_{s1}$ might be observed as 
one approaches much stronger anisotropy regime than those considered in 
this study.  

We like to thank P.~A. Lee, F. Niedermayer, B.~C. Tiburzi, and
U.-J. Wiese for useful discussions and comments on
the manuscript. We also like to thank 
T. Pardini, R.~R.~P. Singh, and O.~P. Sushkov for correspondence and 
providing their
series expansion results in \cite{Pardini08}. The simulations 
in this study were performed using the ALPS library \cite{Troyer08}. 
This work is supported in part by funds provided by the Schweizerischer 
Nationalfonds (SNF).

\end{document}